\documentclass[11pt,a4paper]{emulateapj}
\usepackage{epsfig}
\usepackage{amsmath}
\usepackage{natbib}

\begin{document}

\title{Particle Acceleration in Relativistic Electron-Ion Outflows.}
\author{Nicole M. Lloyd-Ronning$^{1, 2}$ \& Chris L. Fryer$^{1,2}$}
\affiliation{$^1$ CCS-2, Los Alamos National Laboratory, Los Alamos, NM, 87545, USA}
\affiliation{$^2$ Center for Theoretical Astrophysics, Los Alamos National Laboratory, Los Alamos, NM, 87545, USA}
\email{lloyd-ronning@lanl.gov} 
\date{\today}

\begin{abstract}
  We use the Los Alamos VPIC code to investigate particle acceleration in relativistic, unmagnetized, collisionless electron-ion plasmas.  We run our simulations both with a realistic proton-to-electron mass ratio $m_{p}/m_{e} = 1836$, as well as  commonly employed mass ratios of $m_{p}/m_{e} =100$ and $25$, and show that results differ among the different cases. In particular, for the physically accurate mass ratio, electron acceleration occurs efficiently in a narrow region of a few hundred inertial lengths near the flow front, producing a power law $dN/d\gamma \sim \gamma^{-p}$ with $p \sim -2$ developing over a few decades in energy, while acceleration is weak in the region far downstream.  We find  $20\%, 10\%$, and $0.2 \%$ of the total energy given to the electrons for mass ratios of $25$, $100$, and $1836$ respectively at a time of  $2500 w_{p}^{-1}$.  Our simulations also show significant magnetic field generation just ahead of and behind the the flow front, with about $1 \%$ of the total energy going into the magnetic field for a mass ratio of $25$ and $100$, and $0.1 \%$ for a mass ratio of and $1836$. In addition, lower mass ratios show significant fields much further downstream than in the realistic mass ratio case. Our results suggest the region and energetic extent of particle acceleration is directly related to the presence of magnetic field generation.  Our work sheds light on the understanding of particle acceleration and emission in gamma-ray bursts, among other relativistic astrophysical outflows, but also underscores the necessity of optimizing numerical and physical parameters, as well as comparing among PIC codes before firm conclusions are drawn from these types of simulations. 
 \end{abstract}

\section{Introduction}
 
  Highly relativistic outflows occur in a variety of astrophysical objects, with particular relevance to gamma-ray bursts and active galactic nuclei.  Radiation from very high energy particles is observed in these (and indeed many other astrophysical) environments.  For example, gamma-ray bursts emit photons over a wide range of energies, extending to $\sim 100$  GeV (Abdo et al. 2009), with most of the energy emitted in the $\sim 10keV-1MeV$ range during the first 100 seconds or so. Their spectra appear to have both a thermal and non-thermal component, with the latter usually dominant (Guiriec et al. 2013, 2014, 2015; however, see Vurm \& Beloborodov 2015 who argue that the $\gamma$-ray spectrum can take  on a non-thermal shape due to various dissipation processes below the photosphere). This apparent non-thermal spectrum suggests that the underlying energy distribution of radiating particles is a power-law that extends to very high energies $\gamma \geq $ hundreds (see, e.g., Lloyd-Ronning \& Petrosian 2001, for a discussion relating the observed photon spectrum to the underlying particle energy distribution).  Although the emission mechanism of gamma-ray bursts has itself been a long-standing open problem in this field (see Pe`er 2015, for a recent review of GRB emission mechanisms), perhaps an even more fundamental and pressing problem is how the radiating particles achieve a non-thermal distribution - {\em in other words, what is the mechanism by which they are accelerated to such high energies}? And, given that most of the energy is emitted in this regime, how is it that such a high efficiency of particle acceleration is achieved?

 It was realized early on (e.g. Fermi 1949; Krymsky 1977; Bell 1978; Axford et al. 1977; Blandford \& Ostriker 1978) that non-relativistic shocks can be an efficient accelerator of particles via the so-called first order Fermi mechanism, by which particles gain energy from repeated crossings of a shock front.  It is relatively straightforward to show that for an isotropic distribution of electrons in a non-relativistic, plane-parallel shock, a balance between particle crossings and escape leads to a power-law distribution of electron energies $dN/dE \sim E^{-p}$, with an index of $p$ that depends only on the shock compression ratio.  For strong shocks of adiabatic index 5/3 (non-relativistic), this “universal index” is $p=2$. (Baring 1997, and references therein).

 The Fermi mechanism is complicated in the case of relativistic shocks.  Relativistic shocks, with adiabatic index $4/3$, are more compressible than non-relativistic shocks so we would expect a different power-law index for the distribution of their energies.  However, there is not a straightforward relation between the power-law index $p$ and shock compression ratio in the relativistic case.  This is because the underlying assumption of isotropy no longer applies.  Nonetheless, although relativistic shocks move fast enough such that escape can be greater than scattering across the shock front (so that an isotropic distribution is not achieved),  each somewhat rare scattering provides a significant energy boost to a particle compared to the non-relativistic case.  Hence, a power-law for the energy distribution is still expected (Kirk \& Schneider 1987; Ellison et al. 1990), even though a straightforward analytic calculation of this index is lacking.

And, of course, there are other possible mechanisms of acceleration possibly at work in relativistic outflows.  When the plasma Alfven velocity $\beta_{a} = B/(4\pi n mc^{2})^{1/2}$ (in units of the speed of light) is comparable to the shock velocity, particles can be quickly accelerated behind the shock in what is known as the second-order Fermi process or stochastic acceleration (Fermi 1949, 1954; Davis 1956; Schlickeiser 1989; Park \& Petrosian 1995).  Alternatively, it has been recently shown that magnetic reconnection can efficiently accelerate particles in the case of plasmas with high magnetization (Guo et al. 2014, 2015; Liu et al. 2014; Sironi \& Spitkovsky 2014; Kagan et al. 2013).   It is fair to conclude the details of the physics of particle acceleration in many astrophysical contexts is still an open question.  In this paper, we attempt to shed light on this issue by examining acceleration in relativistic, unmagnetized collisionless plasmas, using the Los Alamos VPIC code.

\subsection{Why use Particle-In-Cell codes?}
The basic idea of a PIC code is straightforward:  individual computational particles are tracked in phase space, the currents and charges from these particles are computed, and the resulting fields are calculated directly by solving Maxwell’s equations on a grid.  Each particle's position and momentum is updated via the self-consistently calculated fields.  In the case of a collisionless plasma (applicable to GRBs), we solve the relativistic Vlasov-Maxwell equations:

\medskip

  $  \frac{\partial f_{s}}{\partial t} + c\gamma^{-1}\bar{u} \cdot \bar{\triangledown} f_{s} + \frac{q}{mc} (\bar{E} + c \gamma^{-1} \bar{u} \times \bar{B}) \cdot \bar{\triangledown_{u}} f_{s} = 0$

$ \frac{\partial \bar{B}}{\partial t} = -\bar{\triangledown} \times \bar{E}$

$  \frac{\partial \bar{E}}{\partial t} = \epsilon^{-1} \bar{\triangledown} \times \mu^{-1} \bar{B} - \epsilon^{-1} \bar{J} - \epsilon^{-1} \sigma \bar{E}$

\medskip

\noindent where $f(\bar{r}, \bar{u}, t)$ is the phase-space distribution of a species $s$ with particle charge $q$, and mass $m$,  $c$ is the speed of light in vacuum, $\bar{u}$ is the particle momentum, and $\gamma$ is the relativistic Lorentz factor ($\sqrt{1+u^{2}}$).  The parameters $\bar{E}$ and $\bar{B}$ are the electric and magnetic fields and $\bar{J}$ is the current density, while  $\epsilon$, $\mu$ and $\sigma$ are the background medium permittivity, permeability, and conductivity tensors respectively. 
   PIC simulations sample $f$ with a collection of computational particles, where each computational particle represents a sample of many physical particles.  The PIC simulations solve the particles equation of motion in the presence of electric and magnetic fields:

\medskip

    $\frac{dr}{dt} = c\gamma^{-1} \bar{u}$

    $\frac{du}{dt} = \frac{q}{mc}[\bar{E} + c \gamma^{-1} \bar{u} \times \bar{B}]$

\medskip

 As mentioned above, the electric current is then computed from the particle motion, which is then used to compute the corresponding electric and magnetic fields.
   The details of a PIC simulation lie in how the particles are “advanced”, and corresponding fields computed on the computational grid.  The Los Alamos VPIC code uses an explicit-implicit mixture of leap-frog, Verlet, exponential differencing, and Boris magnetic field rotations. VPIC computes the electric and magnetic fields staggered on the computational grid in an energy-conserving force interpolation scheme.  The fields satisfy a discretized charge continuity relation so charge conservation is upheld.  The details of the particle mover and field computation in VPIC can be found in Bowers et al. 2008, 2009.

  PIC codes have proven to be a powerful computational method to reproduce the behavior of a plasma, particularly for long wavelength modes.   They have reproduced the results in the cases in which analytical work on plasmas is possible.  For example, an unmagnetized plasma with a temperature anisotropy (or, analogously, a counter-streaming flow) is expected to generate large scale magnetic fields due to a Weibel-like instability and, indeed, PIC simulations have shown such fields do arise and are quantitatively in agreement with linear theory (e.g. Kazimura et al. 1998).  Most of this work has been done in the context of non-relativistic or mildly relativistic plasmas. In the case of relativistic plasmas, of interest to gamma-ray bursts and many other astrophysical scenarios, analytic theory is difficult and non-linear effects make such a treatment non-tenable.  Large scale simulations are required to capture the physics of the plasma, and PIC codes are ideally suited to do so.

  Some groups have investigated relativistic plasmas with PIC codes.  The particular problem of magnetic field generation and particle acceleration in the context of gamma-ray bursts has been investigated in certain specialized regimes (e.g., Nishikawa et al. 2005; Hededal et al. 2004; Ramirez-Ruiz et al. 2007; Spitkovsky 2008; Keshet et al. 2009; Nishikawa et al. 2009; Sironi et al. 2013; Choi et al. 2014; Ardaneh et al. 2015).  However, the computational technicalities and approximations made between different PIC codes deem it necessary to make comparisons among groups in order to put the results on firm footing. In particular, many of the PIC simulations of relativistic shocks have been for only a few particles per computational cell, and - more importantly - an unphysical proton-to-electron mass ratio. The advantage of the Los Alamos VPIC code is the code’s ability to handle and improve upon these limiting approaches.

\subsection{Aim of This Paper}
  The main goal of this paper is to present the results of particle acceleration from VPIC simulations of relativistic, unmagnetized, collisionless electron-ion plasmas, for three different  proton-to-electron mass ratios (25, 100, and the physically accurate 1836) and compare the results.  In \S2, we describe our simulation set-up, and the necessity of dealing with numerical Cherenkov radiation in relativistic PIC simulations. In \S 3, we present our results, focusing in particular on the electron energy distribution, the primary radiating particles in gamma-ray bursts.  We show how the results differ for different mass ratios in contrast to previous claims (e.g. Sironi et al. 2013), and that the differences are related to the magnetic field generation/structure in each case.  Discussion and conclusions are presented in \S4 and 5.

\section{Simulation Set-Up}
  
Our simulations are run in two dimensions (the simulation region extends in x and z; however, the particles and fields are tracked in three dimensions), in the reference frame of the far downstream region of the plasma.   In this frame, the contact discontinuity at time = 0 between the upstream and downstream region (on the left boundary in the horizontal x direction) is represented by a reflecting (for the particles), conducting (for the fields) wall.  The right boundary in the x-direction is absorbing for particles and fields, while upper and lower boundaries are periodic.   To compare with similar studies (e.g. Spitkovsky 2008; Sironi et al. 2013), particles are loaded as a Maxwellian with a temperature $T=0.05m_{e}c^{2} = 25keV$ (i.e. a cold plasma), and then boosted with a relativistic Lorentz factor of $\Gamma=15$ toward the reflecting/conducting wall. Note that, here, we examine only the region in front of the contact discontinuity.  After time = 0, we call the region behind the flow front - but in front of the contact discontinuity - ``downstream``.  Our time step is $0.026 w_{po}^{-1} = 0.026\sqrt{\gamma}w_{pr}^{-1}$, where $w_{po} =(4 \pi n e^{2}/m)^{1/2}$ is the non-relativistic plasma frequency, and $w_{pr} = w_{po}/\sqrt{\gamma}$ is the relativistic plasma frequency for particles with internal Lorentz factor $\gamma$ (note that for a cold plasma, as set up in the initial conditions of these simulations, $w_{pr} = w_{po}$).  Our grid scale is $0.06(c/w_{po})$, in order to sufficiently resolve the relativistic Debye length $\lambda_{D} = \lambda_{D_{o}}/\sqrt{\Gamma} \sim 0.05(c/w_{po})$.  We have run test simulations with a grid scale $< \lambda_{D}$ up to $>>\lambda_{D}$ and find the results remain consistent as long as our cell size is smaller than a few $\lambda_{D}$.  Our box size is  $5000 (c/w_{po})$, and the simulations are run for up to $\sim 2500 w_{po}^{-1}$.  We use 4 particles-per-cell per species (8 total), with a proton-to-electron mass ratio of $25, 100,$ and $1836$. We also ran simulations with a similar set up, but with more particles-per-cell (from 20 to 50).  We found that the number of particles per cell does not significantly affect our results, but that varying the proton-to-electron mass ratio can change results markedly.

\subsection{On Numerical Cherenkov}
 A very important issue that arises in any relativistic numerical simulation is the so-called numerical “Cherenkov radiation” (e.g. Greenwood et al. 2004).  In such a scenario, the particle speed can sometimes exceed the numerical speed of light, leading to unphysical, high frequency electromagnetic fields.  What is even worse, these numerical fields can sometimes cause an anisotropic particle heating which can lead to lower frequency modes that appear identical to physical Weibel modes (but are not).  A robust and physically justified way of dealing with this numerical artifact is necessary. 
 
\begin{figure}
 \begin{center}
      \epsfxsize=7.5cm\epsfbox{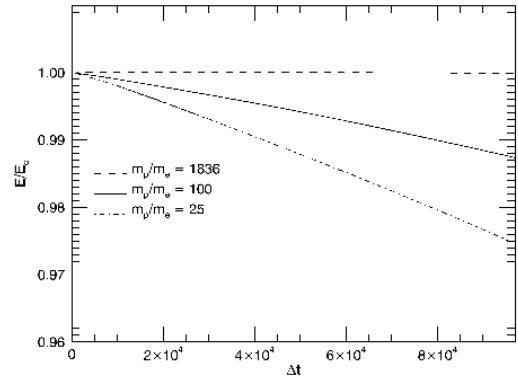}
       \caption[Energy Conservation]{Energy conservation $E/E_{o}$, where $E$ is the total energy and $E_{o}$ is the initial energy in all of the fields and particles for our simulations given different values of the mass ratio $m_{p}/m_{e} = 1836, 100$, and $25$.}
    \label{fig:EnConsAll}
 \end{center}
\end{figure}

  There are several techniques to do so.  For PIC codes that use momentum conserving force-interpolation schemes, a higher order particle shape function is used, which helps mitigate the numerical Cherenkov effect (Meyers et al., 2014; Huang et al., 2015; Eastwood, 1991).  Various filtering schemes are often employed to lessen the numerical grid-scale modes (Greenwood et al., 2004).  Los Alamos’ VPIC code is currently set up to damp the short wavelength/high frequency modes using a “transverse current adjustment” method (Eastwood, 1991; Bowers et al. 2008), which introduces a computationally inexpensive current $\bar{J} = \tau \partial_{t}(\bar{J} - \triangledown \times \mu \bar{B})$ that damps the unphysical, short wavelength, grid-scale modes, using the damping parameter  $d = \Delta t/\tau$.
Of course, we must balance the amount of damping with the preservation of any physical modes that may arise on these scales.  We have attempted to mitigate the numerical Cherenkov through both a highly resolved spatial grid and time step, and employing the damping parameter mentioned above.  We found that a damping parameter $d= 0.3$, suppresses the numerical noise without affecting the physical results of simulations.  We have also checked for artificial numerical heating by monitoring the far upstream particle distribution, and have confirmed that this distribution retains its original form throughout our entire simulation.  Figure~\ref{fig:EnConsAll} shows the global energy conservation in the simulations presented in this paper for the three mass ratios employed and for a damping parameter $d=0.3$.  We note that energy conservation is worse for lower mass ratios and does not appear to converge; however, for the length of our simulations energy is conserved within $1.5 \%$.

\section{Results}
We ran our simulations according the set-up described in \S 2, varying the particles per cell, damping parameter, and - most importantly - the proton-to-electron mass ratio.  As discussed above, varying the particles per cell and damping parameter did not have an appreciable affect on our results.  Therefore, below we present the results for varying the mass ratio alone, and focus on the resultant electron energy distribution (in the context of GRBs, electrons radiate most of the energy, and we are ultimately interested in connecting this to the observed photon distribution).  The mass ratio affects our results, in terms of the density structure of the flow, magnetic field generation, and particle acceleration.  As one might expect, the lower mass ratios give results closer to the case of a pure pair plasma (see, e.g., Spitkovsky 2008b), while the physically realistic mass ratio shows some interesting differences.

\begin{figure}[!t]
\centering
\includegraphics[width=0.45\textwidth]{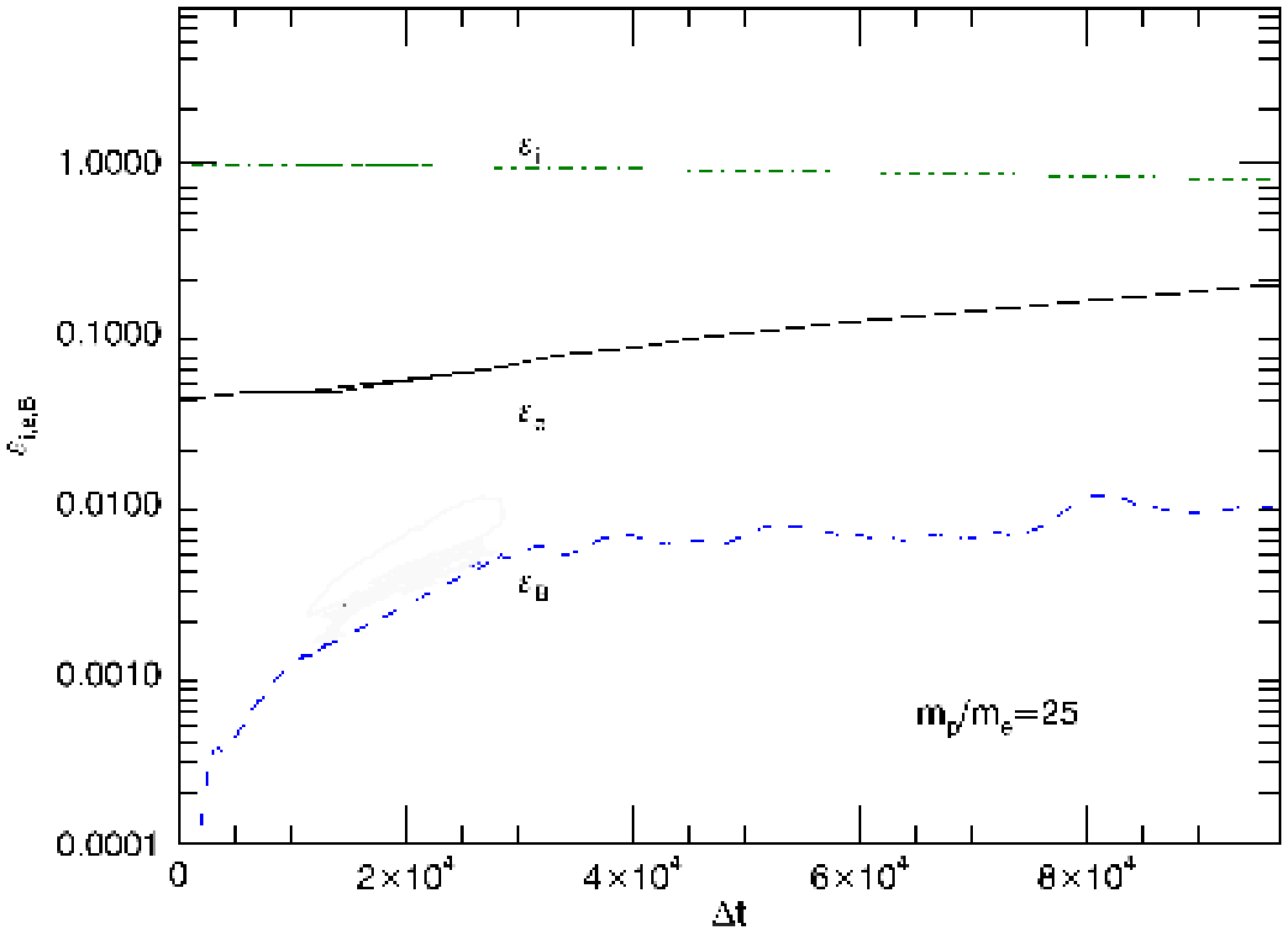}
\caption{Fraction of total energy in the ions, electrons, and magnetic field as a function of simulation time step, for the simulation with $m_{p}/m_{e} = 25$. The simulation ends at $tw_{p} \sim 2500$.
\label{fig:EnpartCo2b}}
\includegraphics[width=0.45\textwidth]{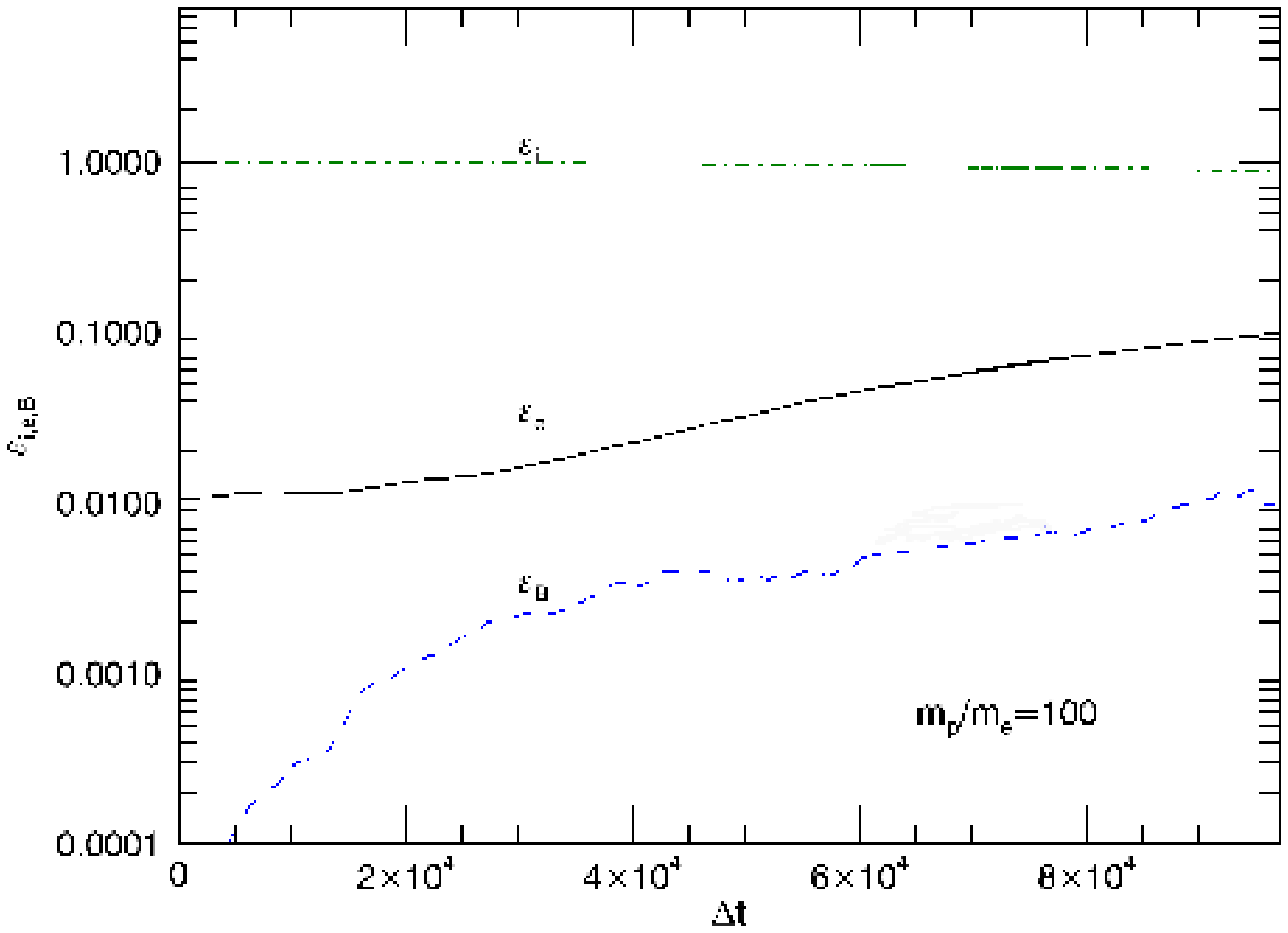}
\caption{Fraction of total energy in the ions, electrons, and magnetic field as a function of simulation time step, for the simulation with $m_{p}/m_{e} = 100$. The simulation ends at $tw_{p} \sim 2500$.
\label{fig:EnpartCo1}}
\includegraphics[width=0.45\textwidth]{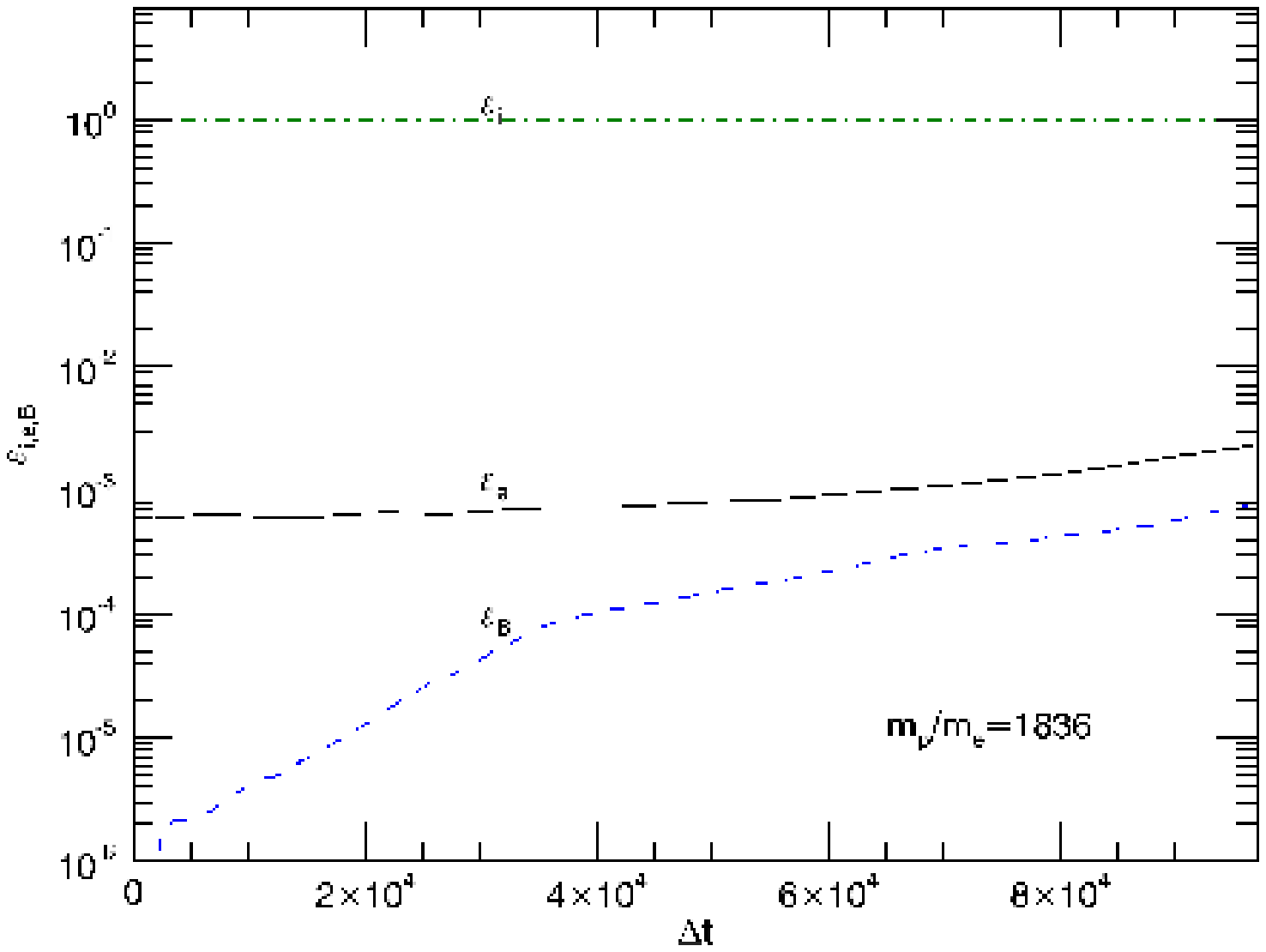}
\caption{Fraction of total energy in the ions, electrons, and magnetic field as a function of simulation time step, for the simulation with $m_{p}/m_{e} = 1836$. The simulation ends at $tw_{p} \sim 2500$.
\label{fig:EnpartCo3}}
\end{figure}

Figures~\ref{fig:EnpartCo2b},\ref{fig:EnpartCo1}, and\ref{fig:EnpartCo3} show the fraction of energy in the ions, electrons and magnetic field for the cases of $m_{p}/m_{e} = 25, 100$ and $1836$ respectively.  In all three cases, the ions slightly lose energy as a function of time in the simulation, as the energy of the electrons and magnetic field increases (note the different scales on the y-axes of the figures).  For the lowest mass ratio, $m_{p}/m_{e} = 25$, we find about $20 \%$ of the energy in the electrons and $1\%$ in the magnetic field by then end of our simulation runs ($tw_{p} \sim 2500$).  For $m_{p}/m_{e} = 100$, these numbers change to about $10\%$ of energy in the electrons and $0.8\%$ in the magnetic field.  Finally, for the realistic mass ratio of $1836$, $0.2\%$ of the energy is in the electrons, while about $0.1\%$ is in the magnetic field.  
These values are in the expected range for GRBs, based on estimates from their spectra and light curves (see, e.g.  Beniamini et al. (2015), who discuss constraints on $\epsilon_{B}$ in particular).

\subsubsection{Magnetic Field and Particle Densities}

\begin{figure}[!t]
\begin{center}
  \includegraphics[width=0.45\textwidth]{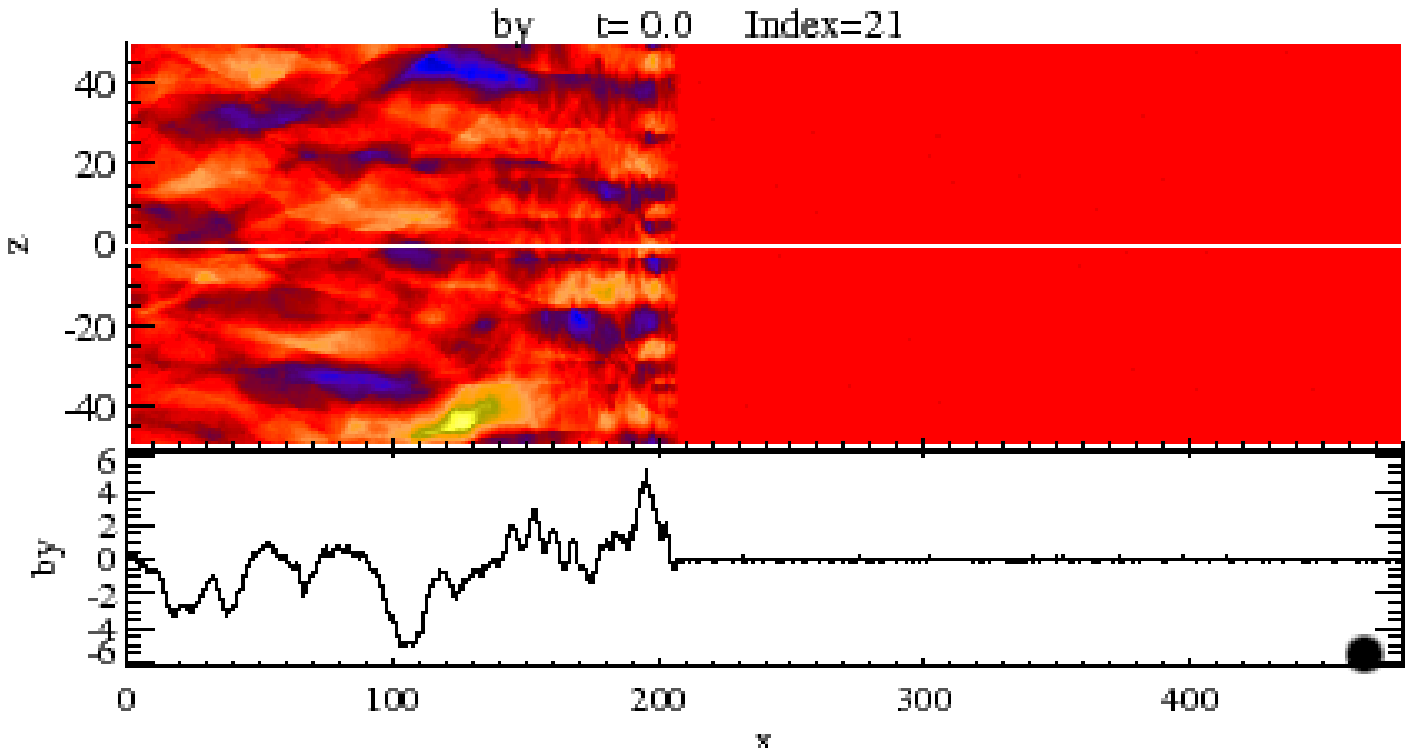}
      \caption[density]{Normalized y-component of the magnetic field as a function of x and z (in units of $c/w_{p}$) for $m_{p}/m_{i}=25$ at $t w_{p} \sim 200$, illustrating the development of the two stream Weibel-like instability in our plasma. The bottom portion of the panel shows a slice through z=0. }
    \label{fig:Bym25}

   \includegraphics[width=0.45\textwidth]{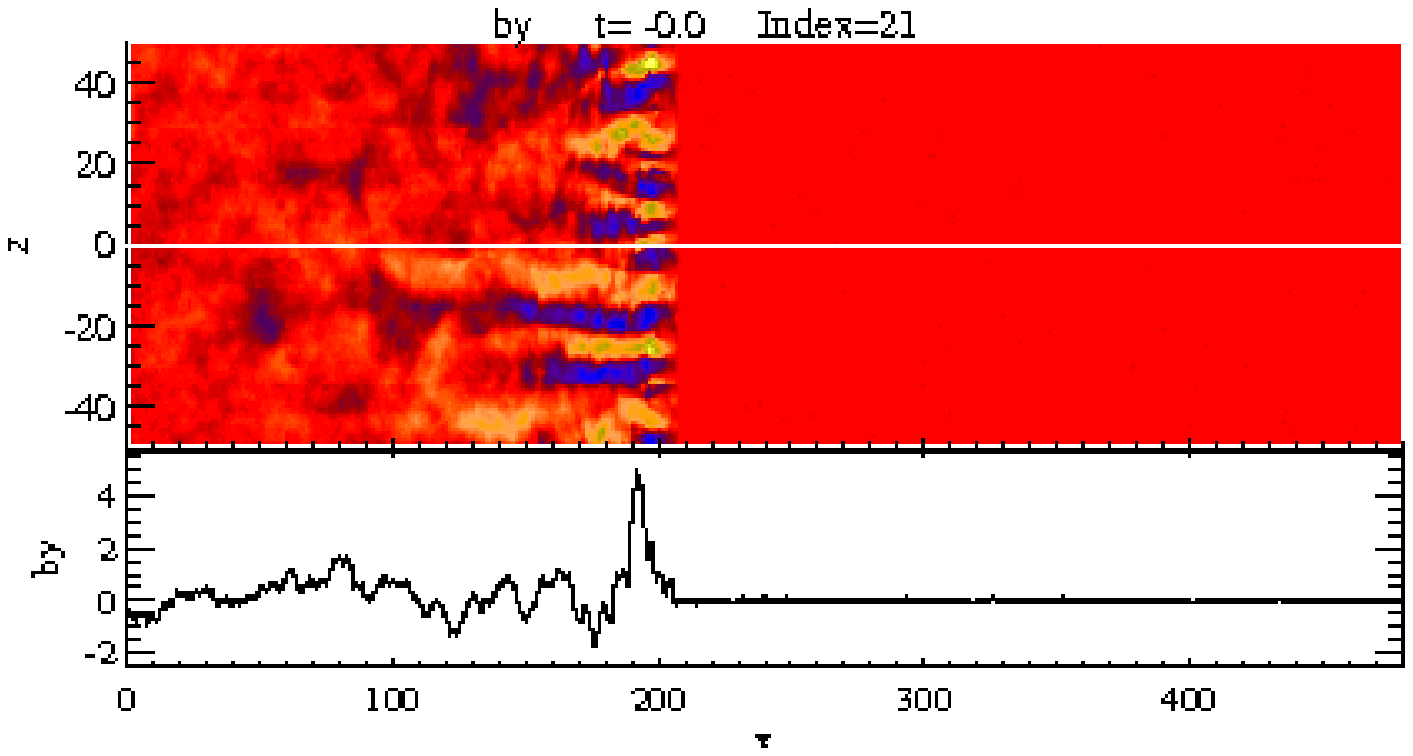}
      \caption[density]{Normalized y-component of the magnetic field as a function of x and z (in units of $c/w_{p}$) for $m_{p}/m_{i}=1836$ at $t w_{p} \sim 200$.  The bottom portion of the panel shows a slice through z=0.  Note that the field is not sustained in the downstream region, in contrast to the lower mass ratio case.}
 \label{fig:Bym1836}
 \end{center}
\end{figure}

\begin{figure}[!t]
\begin{center}
\includegraphics[width=0.45\textwidth]{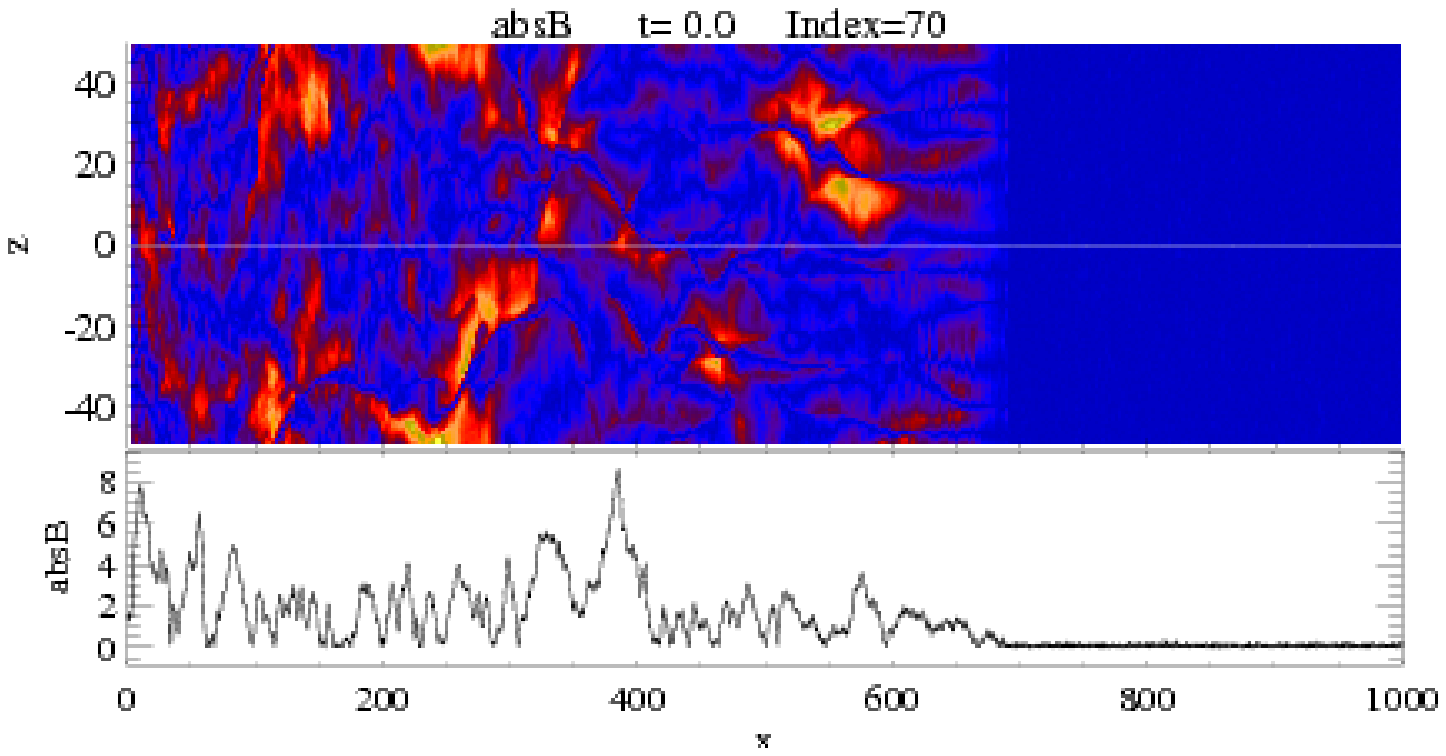}
      \caption[density]{Normalized absolute value of the y-component of the magnetic field as a function of x and z (in units of $c/w_{p}$) for $m_{p}/m_{i}=25$ at $t w_{p} \sim 700$. The bottom portion of the panel shows a slice through z=0. }
\label{fig:Co2babsB70}

\includegraphics[width=0.45\textwidth]{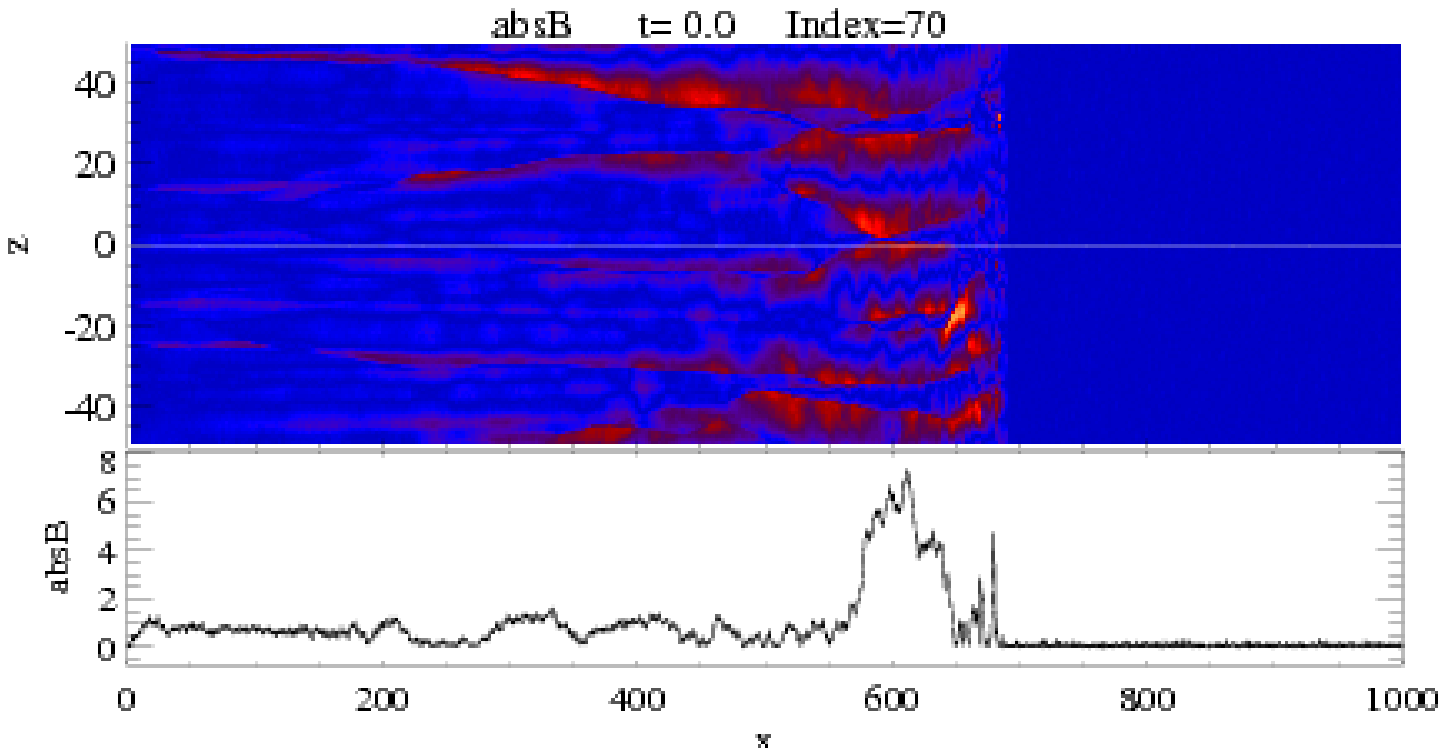}
      \caption[density]{Normalized absolute value of the y-component of the magnetic field as a function of x and z (in units of $c/w_{p}$) for $m_{p}/m_{i}=1836$ at $t w_{p} \sim 700$. The bottom portion of the panel shows a slice through z=0. }
\label{fig:Co3absB70}
\end{center}
\end{figure}

  Figures~\ref{fig:Bym25}  and~\ref{fig:Bym1836} show the y-component of the magnetic field at early times for $m_{p}/m_{e} = 25$ and $1836$ respectively (for brevity, we show the extremes of the mass ratios we used).   The magnetic field develops from a two-stream instability, and is sustained in the downstream region for $m_{p}/m_{e}=25$, whereas it decays much more quickly ($< 100$ inertial lengths) when the physical value of the proton-to-electron mass ratio is used.  Figures~\ref{fig:Co2babsB70}  and~\ref{fig:Co3absB70} show the absolute value of the $y$ component of the magnetic field cases of $m_{p}/m_{e} = 25$ and $1836$ respectively.  This appears to be consistent with Lemoine 2015 who showed analytically that magnetic turbulence is significantly damped downstream, with higher mass ratio plasmas damping at a faster rate than lower mass ratio plasmas.

\begin{figure}[h]
\begin{center}
\includegraphics[width=0.45\textwidth]{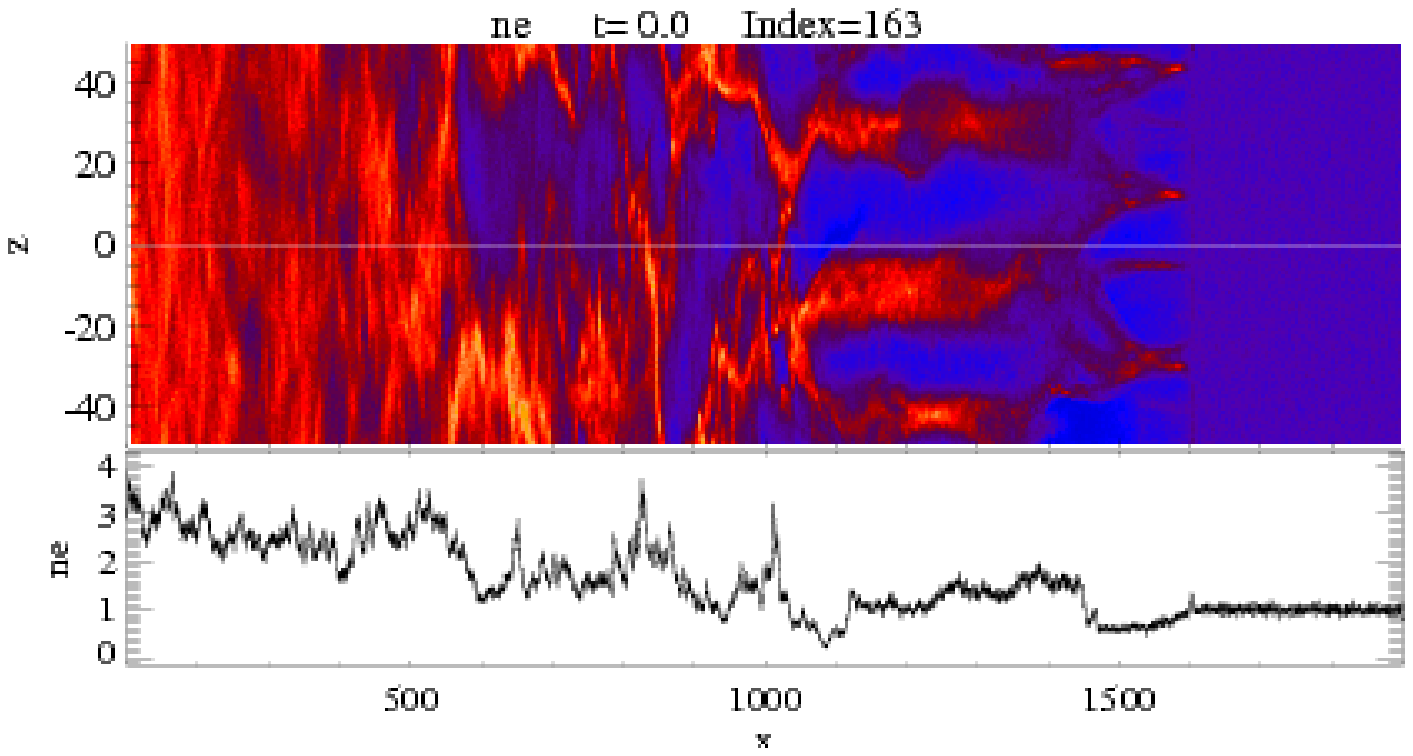}
      \caption[density]{Normalized electron density as a function of x and z (in units of $c/w_{p}$)  for $m_{p}/m_{i}=25$ at $t w_{p} \sim 1600$. }
    \label{fig:Co2bne163}
\end{center}
\end{figure}

\begin{figure}[h]
\begin{center}
\includegraphics[width=0.45\textwidth]{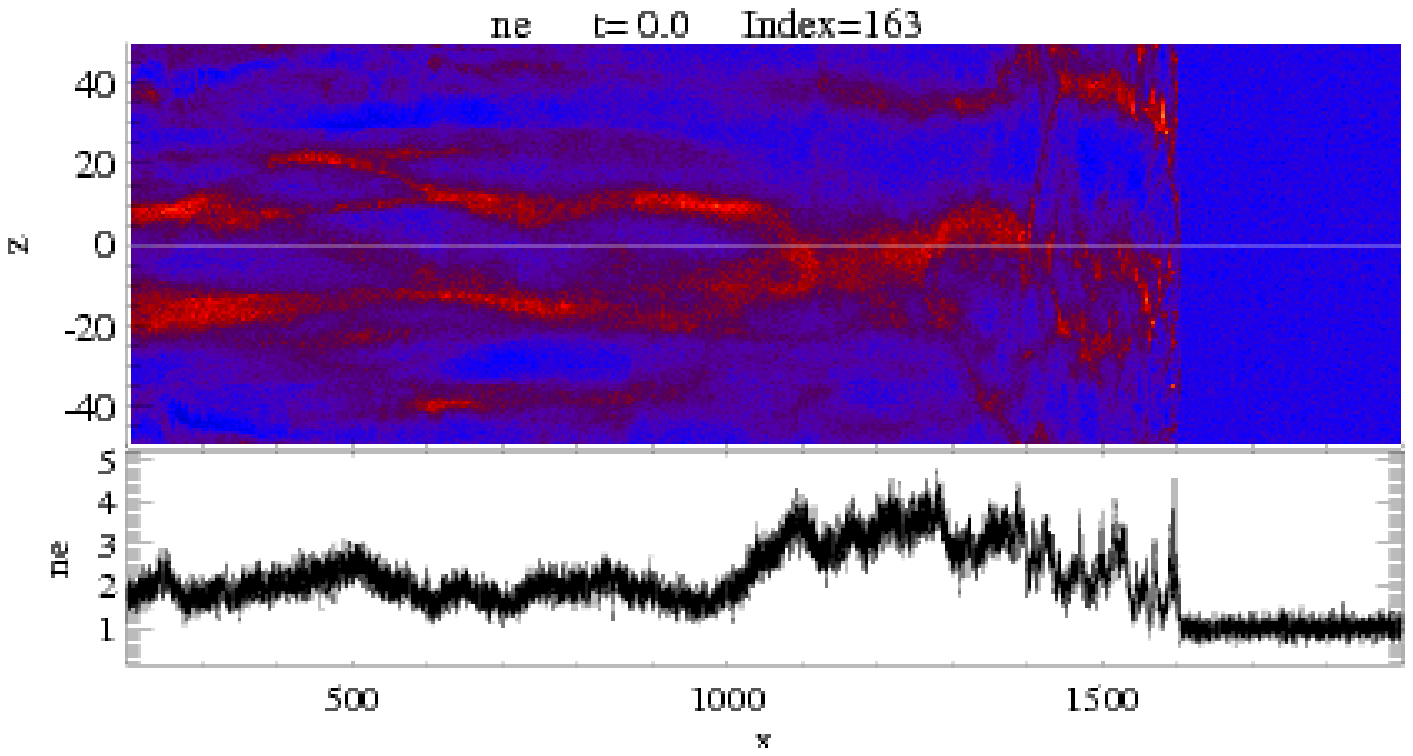}
      \caption[density]{Normalized electron density as a function of x and z (in units of $c/w_{p}$) for $m_{p}/m_{i}=1836$ at $t w_{p} \sim 1600$. }
    \label{fig:Co3ne163}
\end{center}
\end{figure}

\begin{figure}[h]
\begin{center}
\includegraphics[width=0.45\textwidth]{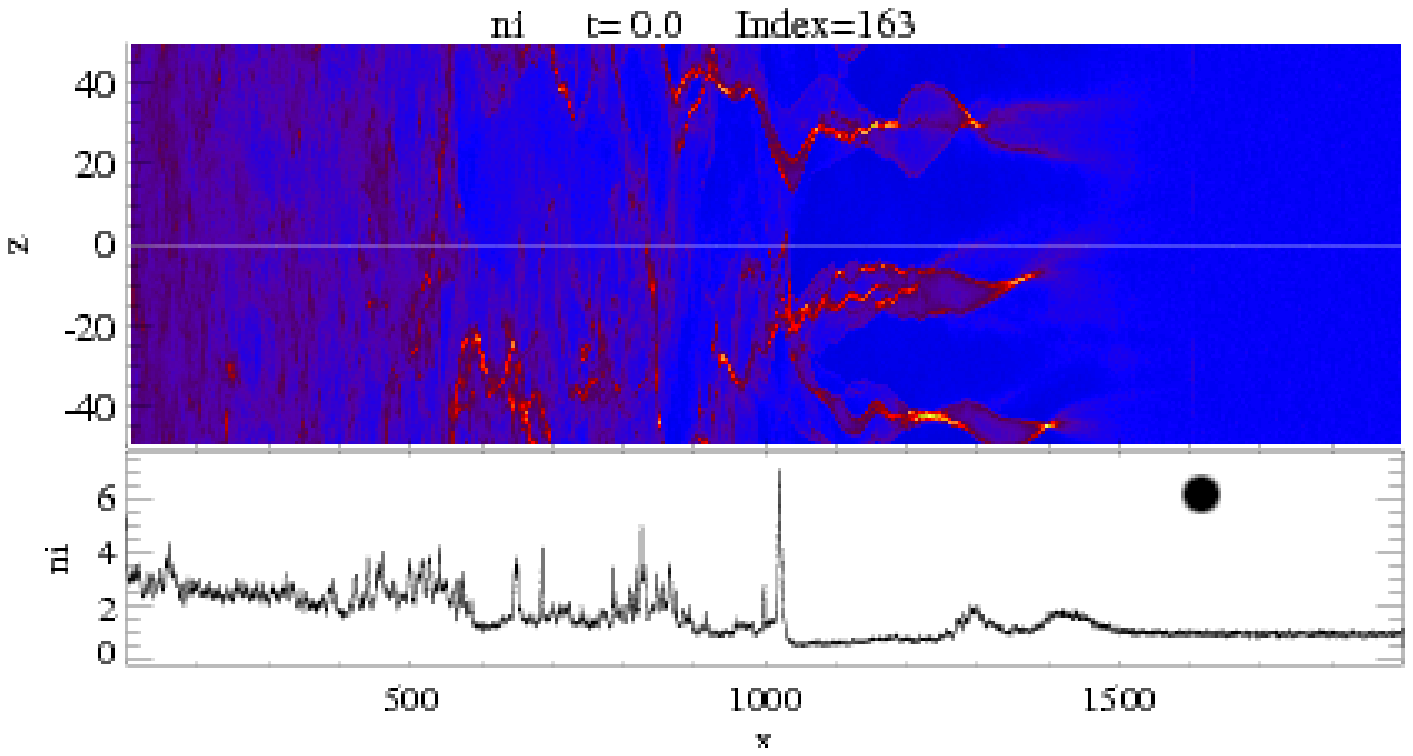}
      \caption[density]{Normalized Ion density as a function of x and z (in units of $c/w_{p}$) for $m_{p}/m_{i}=25$ at $t w_{p} \sim 1600$.}
    \label{fig:Co2bni163}
\end{center}
\end{figure}

\begin{figure}[h]
\begin{center}
\includegraphics[width=0.45\textwidth]{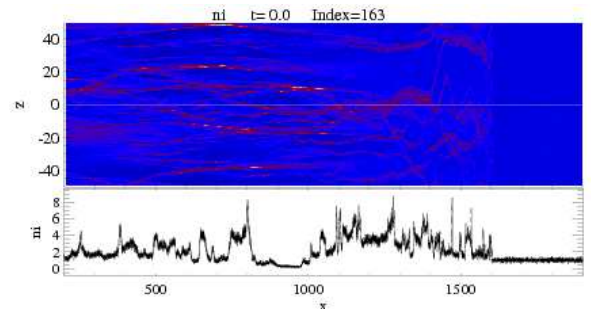}
      \caption[density]{Normalized ion density  (in units of $c/w_{p}$) for $m_{p}/m_{i}=1836$ at $t w_{p} \sim 1600$.}
    \label{fig:Co3ni163}
\end{center}
\end{figure}

  The electron density (normalized to the upstream density) is shown in figures~\ref{fig:Co2bne163} and~\ref{fig:Co3ne163} for the lower and realistic mass ratios respectively, while figures~\ref{fig:Co2bni163} and~\ref{fig:Co3ni163} show the ion density (again normalized to the upstream value) at a representative time of $tw_{p} \sim 1600$.  The lower mass ratio plasmas show a shock developing with an average value of the ratio of downstream to upstream density of $\sim 3$ as expected from 2D relativistic shock jump conditions (Spitkovsky 2008).   The realistic mass ratio case shows a narrow shock ($\sim 500$ inertial lengths) just behind the flow front for the electron density, and significant density structure for the ions.  Our most significant result here is that the shock in the electron density does not extend far downstream for a realistic mass ratio, but does when a lower mass ratio is employed.  We will discuss the implications of this below.
 


\subsubsection{Particle Energy Spectrum}

\begin{figure}[!t]
 \begin{center}
      \epsfxsize=7.5cm\epsfbox{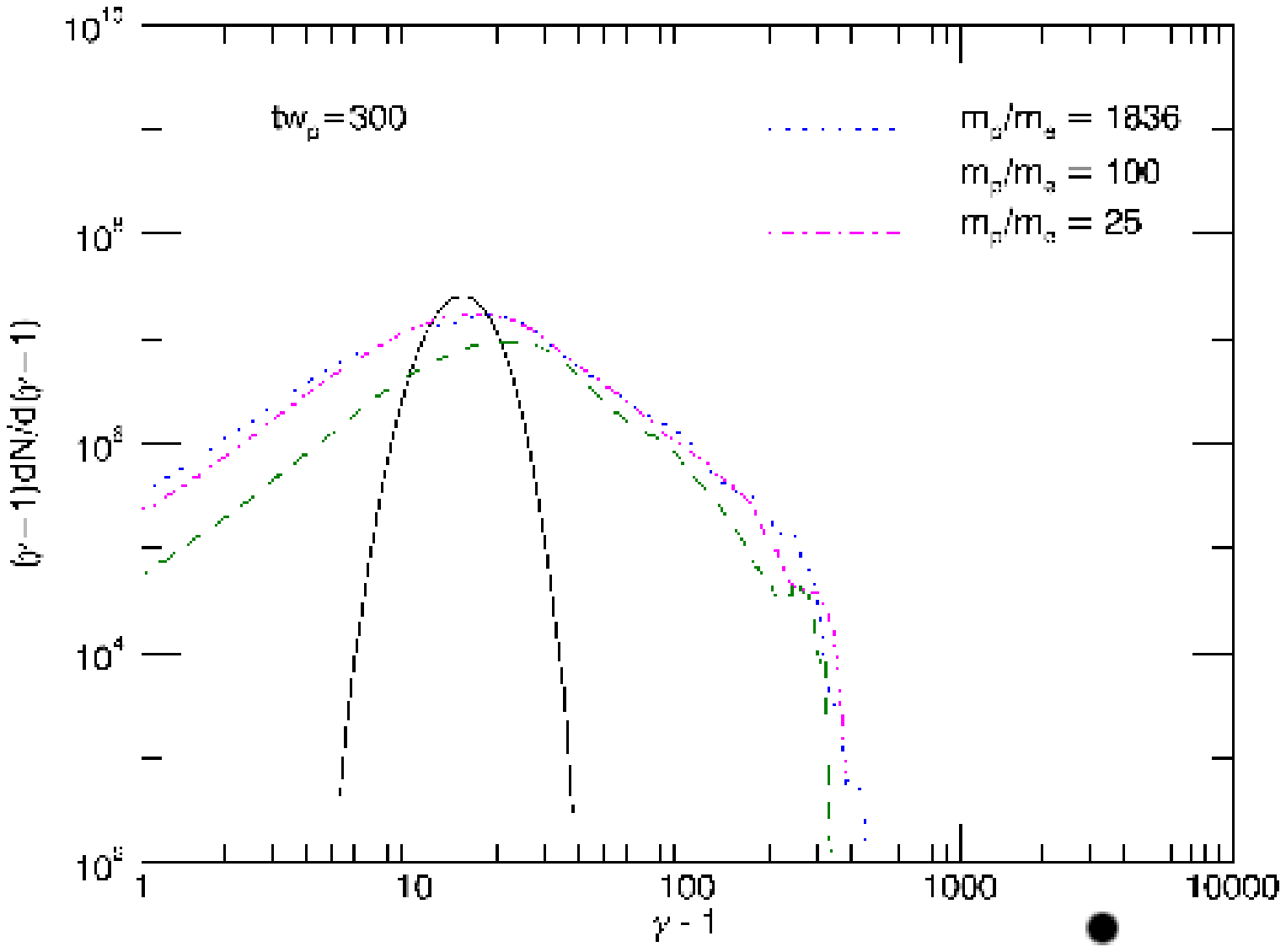}
       \caption[t=300]{Electron energy spectrum in a $200 c/w_{p}$ slice around the shock front at $t w_{p} = 300$, for the three different mass ratios.}
    \label{fig:ElSpecearly}
 \end{center}
 \begin{center}
      \epsfxsize=7.5cm\epsfbox{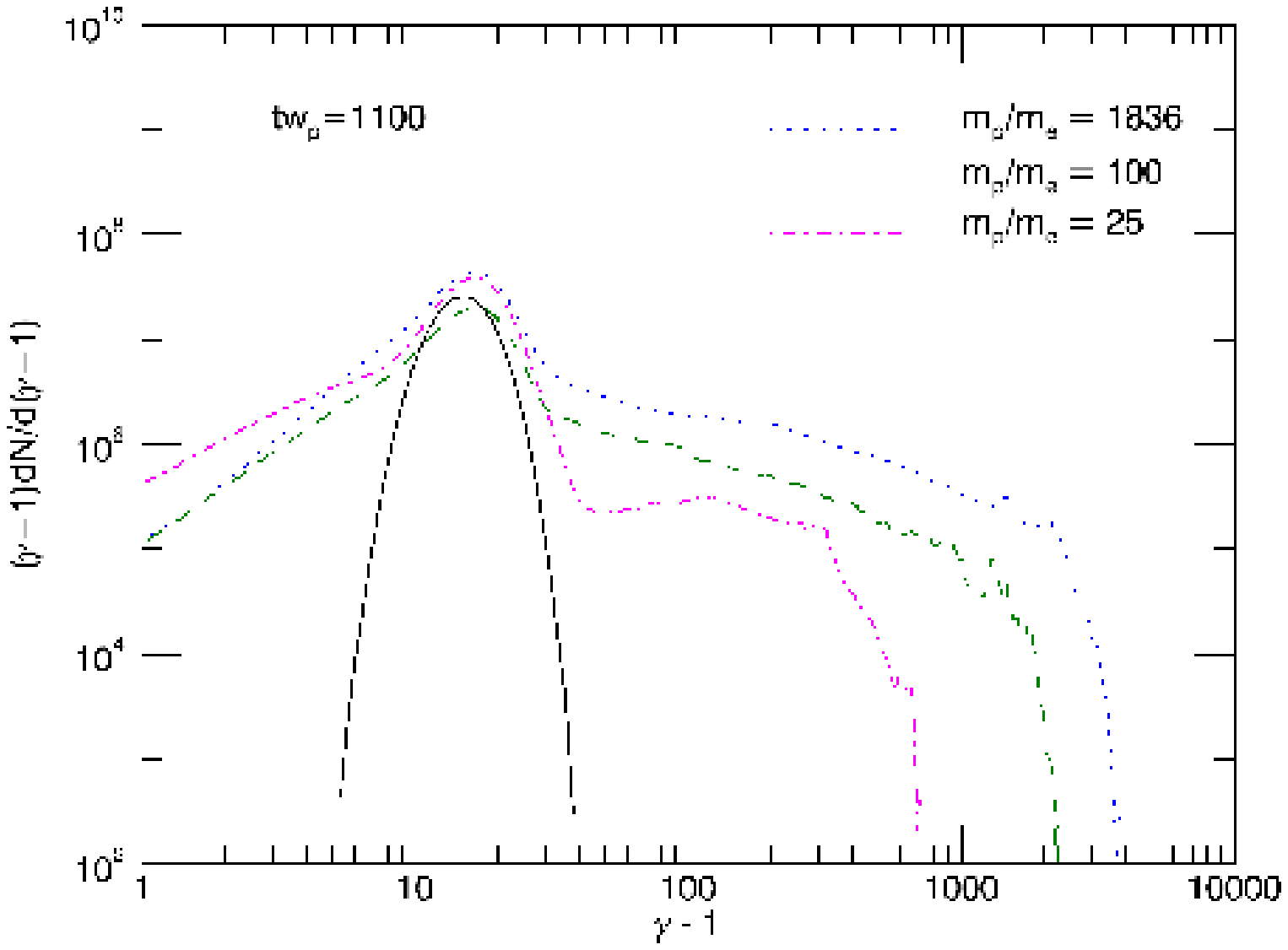}
       \caption[t=1000]{Electron energy spectrum in a $200 c/w_{p}$ slice around the shock front at $t w_{p} = 1100$, for the three different mass ratios.}
    \label{fig:ElSpecmid}
 \end{center}
 \begin{center}
      \epsfxsize=7.5cm\epsfbox{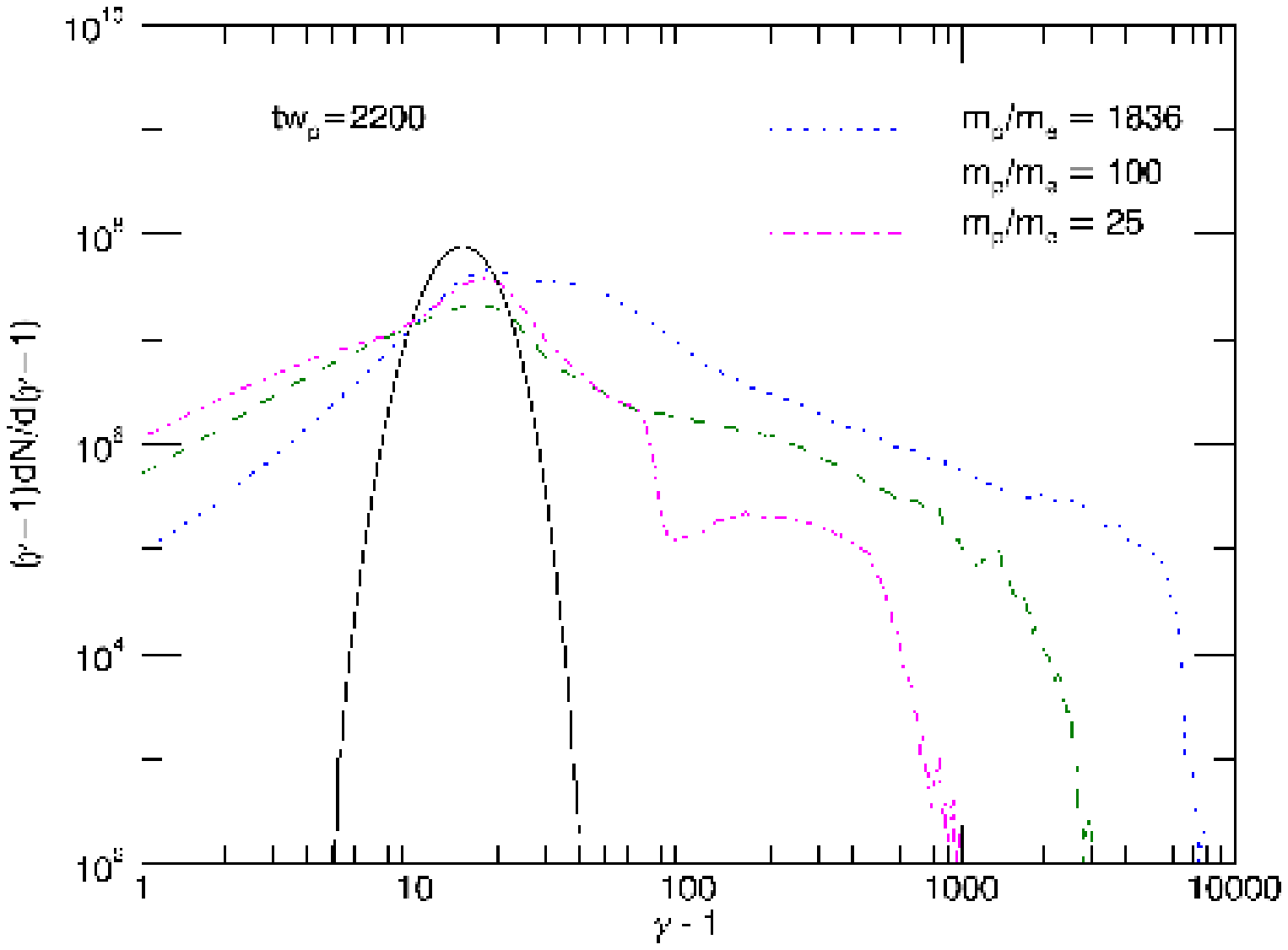}
       \caption[t=1500]{Electron energy spectrum in a $200 c/w_{p}$ slice around the shock front at $t w_{p} = 2200$, for the three different mass ratios.}
    \label{fig:ElSpeclate}
 \end{center}
\end{figure}

 Figures~\ref{fig:ElSpecearly},~\ref{fig:ElSpecmid}, and \ref{fig:ElSpeclate} show the electron kinetic energy distribution $(\gamma-1)dN/d\gamma$ vs. $(\gamma - 1)$  for the three mass ratios we employed in a $200c/w_{p}$ slice in $x$ at three different times.  Figure~\ref{fig:ElSpecearly} shows the particle energy spectrum at a time $t w_{p} \simeq 300$, in a region about $300$ inertial lengths from the contact discontinuity (left boundary).  At this time, the flow front is in the slice we are examining, and all three particle energy spectra are similar, with a power-law developing off the initial thermal peak.  Figures \ref{fig:ElSpecmid} and \ref{fig:ElSpeclate} show the particle energy distribution at $t w_{p} =  1000$ and $2200$ respectively, again in a $200 c/w_{p}$ wide region around the shock front.    One can see the results of the particle energy spectra for the different mass ratios diverge with time, with the higher mass ratio giving rise to higher energy electrons and an extended power-law.  The lower mass ratios show a secondary high energy bump, with maximum energies that remain below that of the realistic mass ratio case.  The magenta dashed line in the last figure (mass ratio of 25), can be compared to Figure 11 of Sironi et al. (2013) who find a similar spectrum for the electrons at this time.     At later times, we find the maximum electron Lorentz factor $\gamma_{max} \propto (t w_{p})^{\sim 0.5}$ for $m_{p}/m_{e} = 25$ and $100$, consistent with the scaling found in Sironi et al. (2013). The realistic mass ratio of $1836$ shows a scaling $\gamma_{max} \propto (t w_{p})^{\sim 1}$, consistent with a classic Bohm scaling law $\propto t^{1}$.

\begin{figure}[!t]
\begin{center}
     \epsfxsize=7.5cm\epsfbox{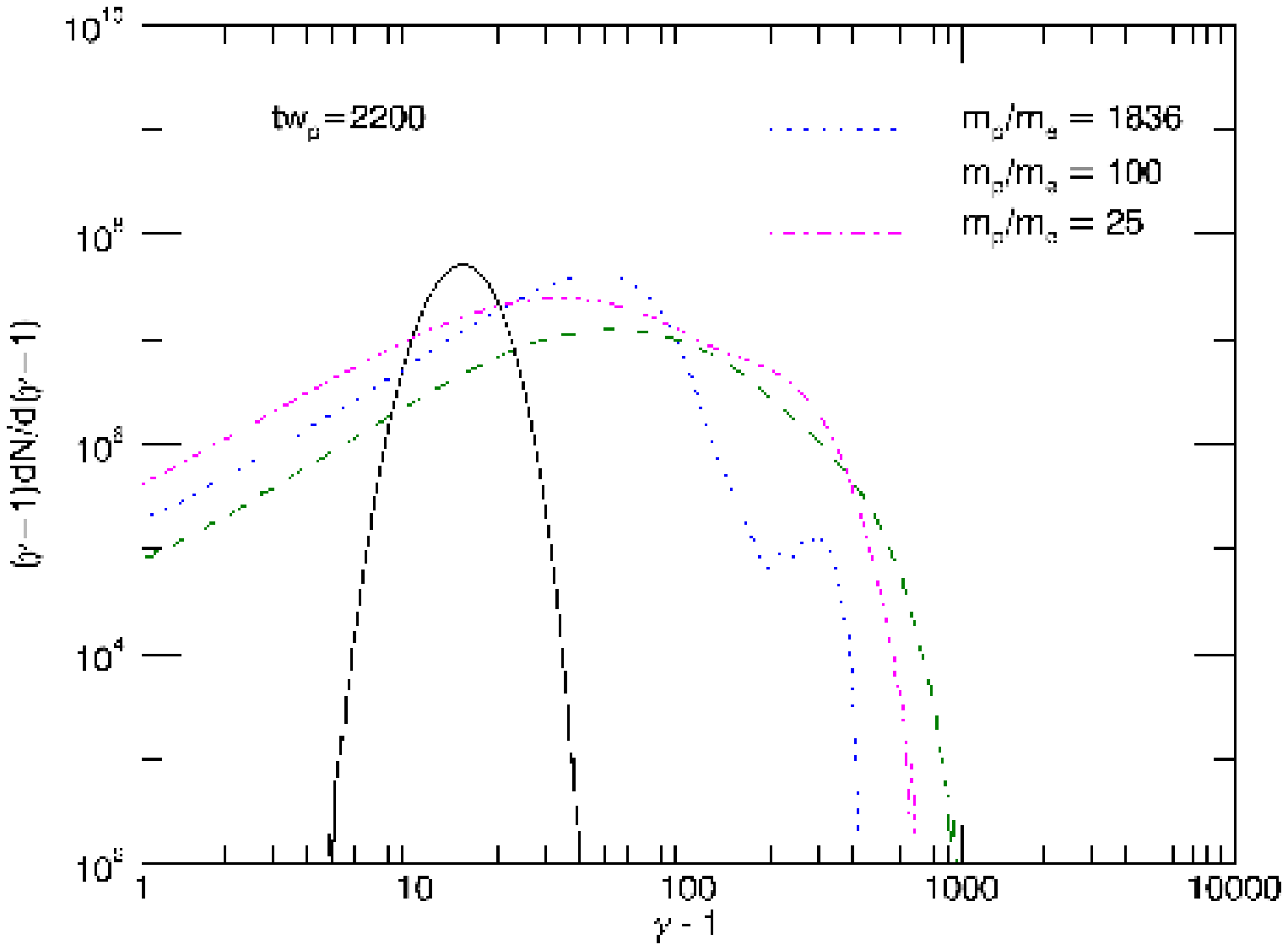}
      \caption[ ]{Electron energy spectrum $\sim 1000$ inertial lengths downstream from the flow front, at $t w_{p} = 2200$, for the three different mass ratios.}
    \label{fig:ElSpecds2}
    \label{fig:ElSpecALLa}
      \epsfxsize=7.5cm\epsfbox{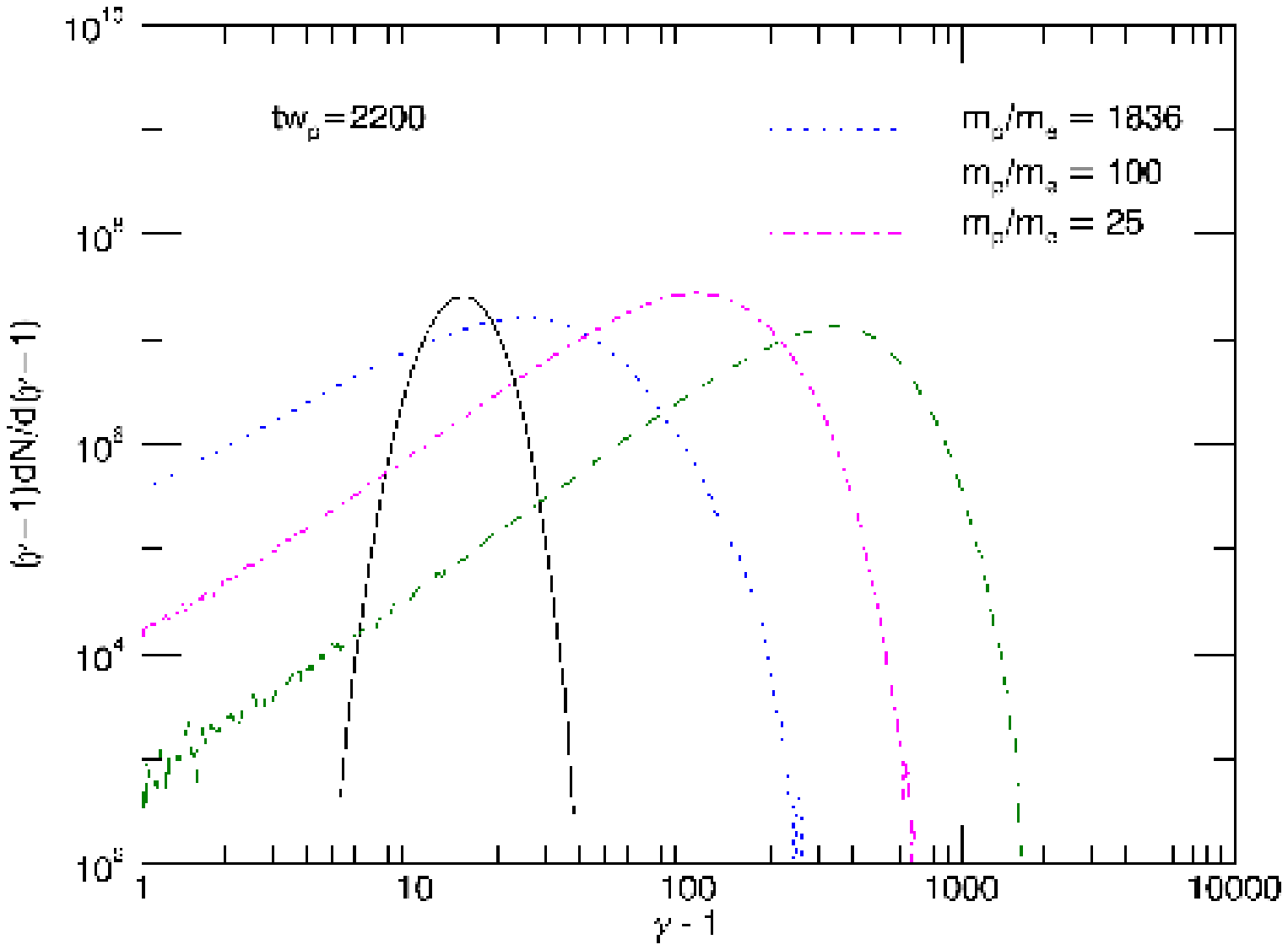}
       \caption[]{Electron energy spectrum $\sim 2000$ inertial lengths downstream from the flow front, at $t w_{p} = 2200$, for the three different mass ratios.}
    \label{fig:ElSpecds3}
 \end{center}
\end{figure}

  Figures~\ref{fig:ElSpecds2} and~\ref{fig:ElSpecds3} show the electron energy distribution for a $200 c/w_{p}$ slice $1000$, and $2000$ inertial lengths {\em downstream}, respectively, from the flow front at $t w_{p} = 2200$.  Within $1000$ inertial lengths of the front, the three spectra overlap, but as we look further downstream, only the lower mass ratios show acceleration in this region. {\em  In other words, acceleration occurs for the realistic mass ratio in only a narrow ($\sim$ hundreds of inertial lengths) region around the shock front, while lower mass ratios show acceleration over a much extended region of thousands of inertial lengths.}  The region of acceleration directly corresponds to both the presence of a shock and extent of magnetic field generation among the different mass ratios. 
 
\begin{figure}[!t]
 \begin{center}
      \epsfxsize=7.5cm\epsfbox{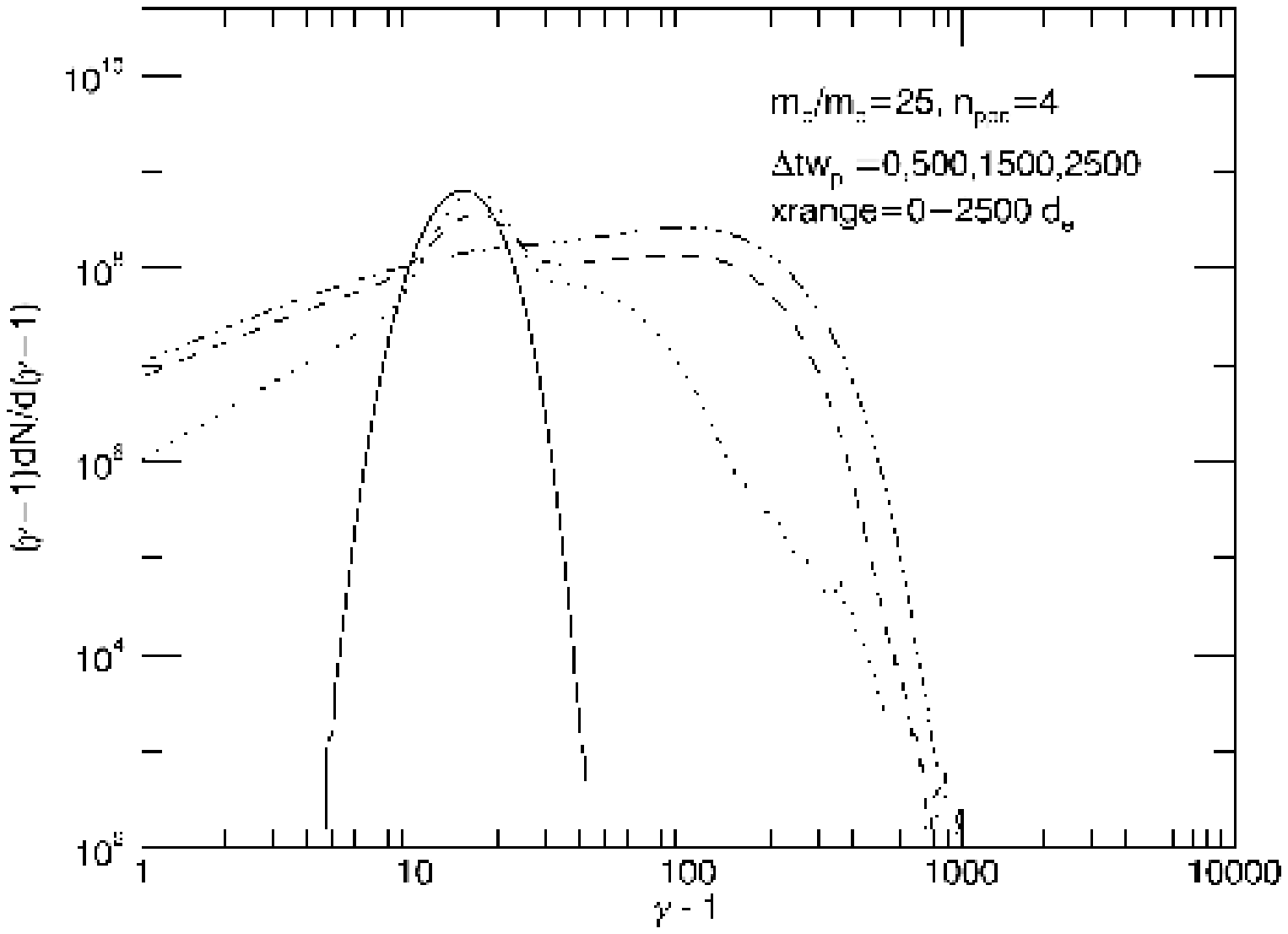}
       \caption[ ]{Electron energy spectrum over the entire simulation region, at $t w_{p} = 0, 500, 1500, 2500$ for a mass ratio $m_{p}/m_{e} = 25$.}
    \label{fig:ElSpecCo2b}
    \epsfxsize=7.5cm\epsfbox{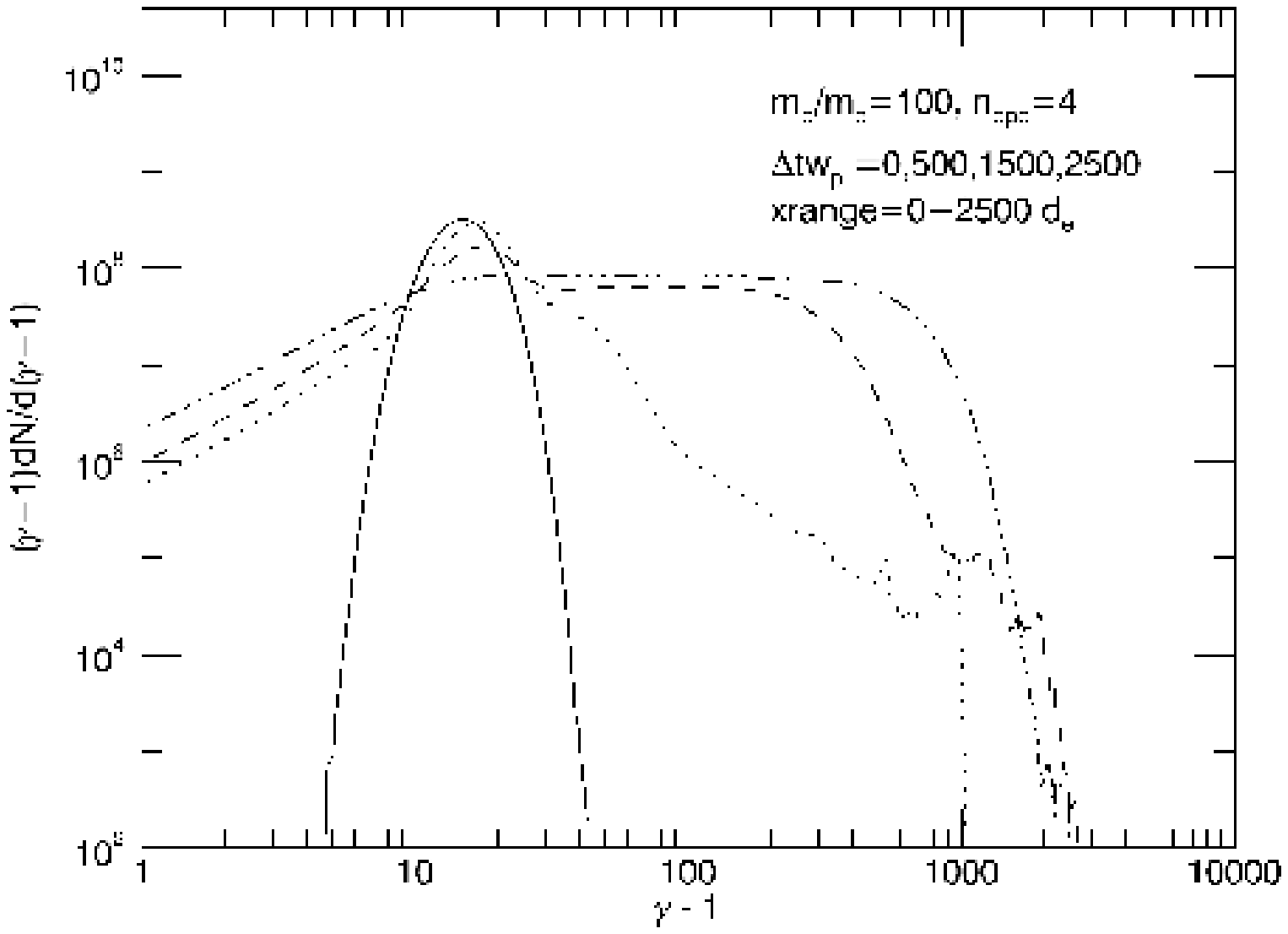}
       \caption[]{Electron energy spectrum over the entire simulation region, at $t w_{p} = 0, 500, 1500, 2500$ for a mass ratio $m_{p}/m_{e} = 100$.}
    \label{fig:ElSpecCo1}
      \epsfxsize=7.5cm\epsfbox{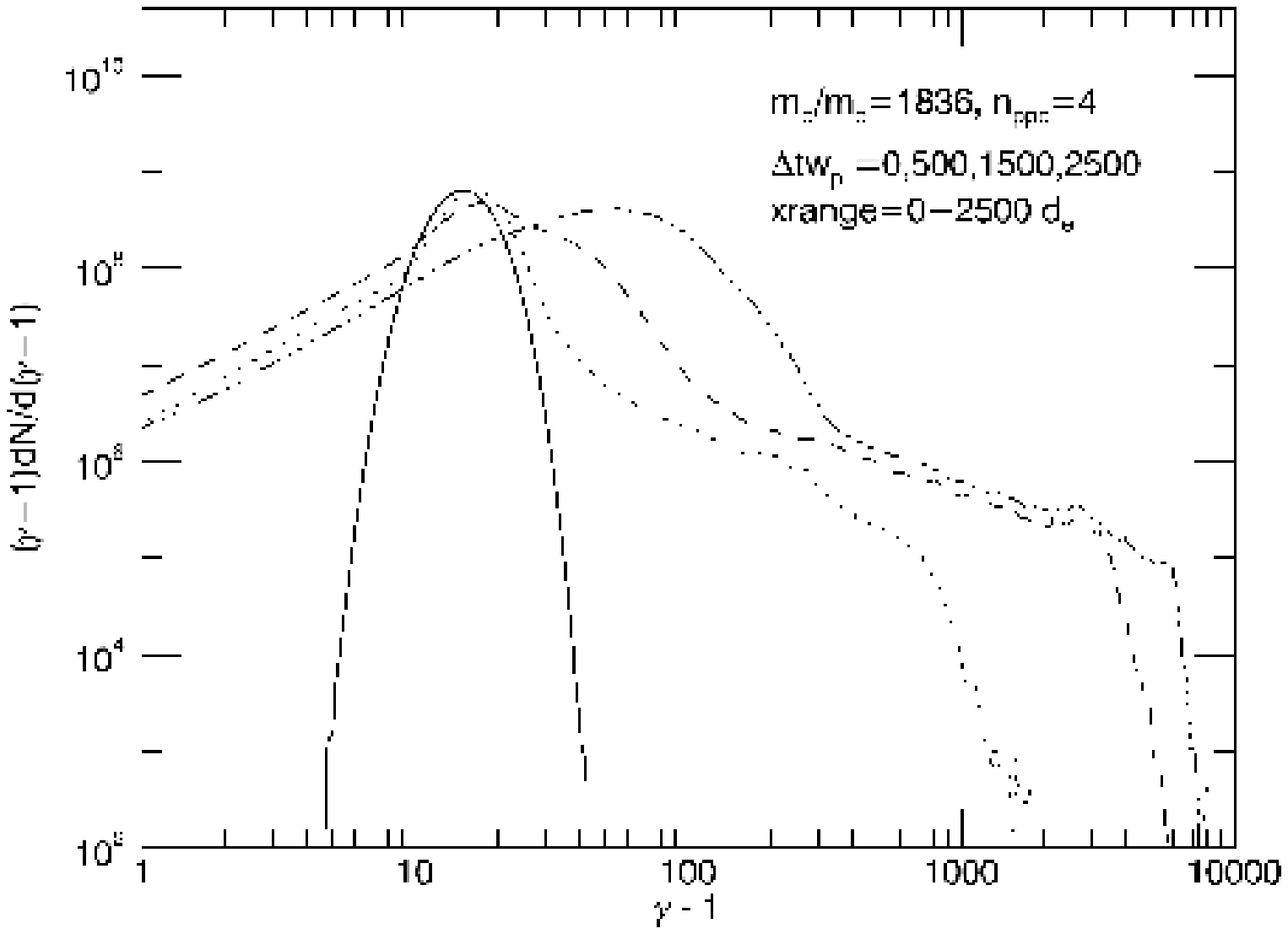}
       \caption[]{Electron energy spectrum over the entire simulation region, at $t w_{p} = 0, 500, 1500, 2500$ for a mass ratio $m_{p}/m_{e} = 1836$.}
    \label{fig:ElSpecCo3}
 \end{center}
\end{figure}

Finally, Figures~\ref{fig:ElSpecCo2b},~\ref{fig:ElSpecCo1}, and~\ref{fig:ElSpecCo3} show the particle energy distribution evolution (for times $t w_{p} = 0, 500, 1500, 2500$) over the entire simulation box, for the mass ratios of $25$, $100$, and $1836$ respectively.   For the lower mass ratio cases, we see a steep power-law at early times ($\Delta t w_{p} \sim 500$); at a later times, we find a very flat, hard spectrum off the initial thermal peak out to some high lorentz factor ($\gamma_{e} \sim 200$ for the mass ratio of 25 and $\gamma \sim 400$ for the mass ratio of 100), and then a sharp drop-off after that.  For the realistic mass ratio case, we see simply an extended power-law of index $\sim 2$ (in $dN/d\gamma$) develop off the initial thermal peak, extending over almost 2 decades in energy at late times. 
   The difference in the shapes of the particle energy spectra for different mass ratios can be attributed to the differences in the presence of the magnetic field.  As we discussed above, the realistic mass ratio case shows strong field generation at the flow front and rapid (a few hundred inertial lengths) decay downstream, whereas the lower mass ratio cases show significant field presence downstream.   This suggests the magnetic turbulence as the source of particle acceleration.  This is why we find acceleration downstream in the lower mass ratio cases, but not in the realistic mass ratio case.


\section{Discussion}
  The differences in our results - particularly the shape of the electron energy distribution - when employing different mass ratios in our PIC simulations can ultimately be attributed to the details of the magnetic field generation.  There are two factors that each come into play - the ability of the plasma to sustain a magnetic field in the downstream region, and the size of the magnetic filaments relative to the particle gyroradius.  Our simulations show that electrons are accelerated in regions of significant magnetic field, generated from a two-stream (Weibel-like) instability mediated by the electrons (indeed analytic calculations easily show the electron Weibel instability grows much faster than the ion Weibel instability, since the growth rate is proportional to the plasma frequency which in turn is proportional to the inverse square root of the mass; see e.g. Stockem Novo et al. 2015).  This strongly suggests that scattering off the magnetic turbulence plays a primary role in the electron acceleration.  However, this scattering will only be efficient when particle gyro-radius ($=mv_{\perp}/eB$, where $m$ is the particle mass, $v_{\perp}$ is the component of velocity perpendicular to the magnetic field, $e$ is the particle charge, and $B$ is the magnitude of the magnetic field) is of order the typical length scale of the magnetic filaments.  These filaments grow in time to a size $\leq \sqrt{\Gamma/2} c/w_{pe}$ (Stockem Novo et al. 2015) when the magnetic field saturates, and are therefore of the size or bigger than the electron gyroradius $r_{g,e} \sim B_{PIC}^{-1} c/w_{pe}$, where $B_{PIC}$ is the normalized value of the magnetic field in our simulations.  This implies the magnetic turbulence is able to scatter electrons effectively, which allows them to cross the shock and gain energy. \footnote{ For small mass ratios, the magnetic field will also scatter the ions (with gyro-radii $r_{g,i} = m_{p}/m_{e} r_{g,e}$  - i.e. smaller mass ratios have smaller ion gyro-radii), and the energy is shared between electrons and ions as born out by the simulations.  For realistic mass ratios, the ion gyro-radius is much larger than the magnetic filaments and therefore the ions are unable to be scattered and accelerated. } And finally, as mentioned above, because the magnetic field decays downstream in the realistic case (but not for lower mass ratios), there is no acceleration downstream in contrast to lower mass ratios in which a field can be sustained downstream.   We plan to explore the details of the acceleration process using a particle tracking method, which we defer to a future paper.

\section{Conclusions}
 In this paper we have investigated particle acceleration in relativistic flows ($\Gamma=15$) for electron-ion plasmas for three different proton-to-electron mass ratios ($25, 100, \& 1836$), using the Los Alamos VPIC code.  We find a marked difference in our results when we change the mass ratio.   Our main conclusions are:
\begin{enumerate}
\item{The magnetic field decays quickly (within a few hundred inertial lengths) downstream for a realistic mass ratio, while it is sustained much further downstream for lower mass ratios (over thousands of inertial lengths).  For the realistic mass ratio case, the region of magnetic field generation, therefore, is $\sim 500 c/w_{p} \sim 10^{8}$ cm $n_{1}^{-1/2}$, where $n_{1}$ is the density normalized to units of $1 cm^{-3}$.  The typical radius of internal shocks is expected to be $\sim 10^{12} cm$ (e.g. Gehrels et al. 2009).  Hence, the region of field generation is extremely narrow compared to radius at which the shocks occur.  This has important implications for the interpretation of the prompt emission as we discuss below. The total amount of energy in the magnetic field at $tw_{p} \sim 2500$ is $\sim 1\%$, $1 \%$, and $0.1\%$ for mass ratios of 25, 100 and 1836 respectively.}
\item{The electron energy distribution varies significantly at late times ($tw_{p} > 500$) between simulations with different proton-to-electron mass ratios.  For a realistic mass ratio, we find particles accelerated to higher energies for higher mass ratios, with a power-law of index $-2$ (in $dN/d\gamma$) developing over a couple of decades in energy.  The lower mass ratio simulations show a flat, hard spectrum off the initial thermal peak and then a sharp decline.}
\item{In the case of a realistic mass ratio, particles appear to be accelerated primarily at the flow front in a fairly narrow region $\sim$ hundreds of inertial lengths, which directly coincides with the region of magnetic field generation.  The particle energy spectrum downstream from the shock shows little acceleration for the realistic mass ratio case, and therefore acceleration can be attributed to the presence of magnetic turbulence.}
\end{enumerate}
Ions are not significantly accelerated in our simulations, with less acceleration for higher proton-to-electron mass ratios.  However, we have only run the simulations for a few hundred inverse {\em ion} plasma frequencies.  In a future publication, we plan to explore the behavior of the ion spectrum in more detail.  If indeed there is little acceleration in the case of a realistic proton-to-electron mass ratio, this has implications for GRBs as the source of high energy neutrinos.

\medskip

 The determination of both the magnetic field and particle energy distribution in relativistic shocks are crucial in attempting to understand the high energy emission from astrophysical environments. For example, for an optically thin synchrotron photon spectrum from an isotropic particle distribution with index $p$, we will observe a break in the spectrum that is proportional to the square characteristic electron energy $\gamma_{e}^{2}$ and the amplitude of the magnetic field. The observed photon spectrum above this break is $\propto -(p+2)/2$ if particles lose all of their energy to radiation quickly and $-(p+1)/2$ if particles cool slowly or in a regime in which they are quickly re-accelerated.  Alternatively, if the particle energy power-law extends to very low energies (i.e. $\gamma_{min}$ is below the detector range), the low energy spectral index $\alpha$ will be related to the power law index $p$ in the same way as described for $\beta$ above.  Hence, the underlying particle distribution and magnetic field have significant consequences for the observed photon spectrum.

Our results show that particles can be accelerated to very high energies, with varying behavior above the thermal peak, and an asymptotic  power-law of index $-2$ (in $ dN/d\gamma)$ extending over about almost two decades in energy, when a realistic proton-to-electron mass ratio is employed.  This is consistent with observations of gamma-ray bursts which show a wide range of high energy behavior (Guiriec et al. 2014, 2015).  Our results also show that the region of high magnetic field and particle acceleration is fairly narrow ($\sim$ hundreds $c/w_{p} = 10^{8}$ cm $n_{1}^{-1/2}$; we note this was also suggested in Hededal et al., 2004).  This has implications for interpreting GRB emission, justifying assumptions of a thin emission region (Mesler et al. 2012, 2014), and weakening the argument for ‘’smearing’’ of features in the light curve due to the radial extent of the emission (e.g. Gat et al. 2013; note, however, that angular effects must still be included and will contribute to the smearing).

Finally, our results have implications for choices of microphysical parameters (e.g. $\epsilon_{e}$ and $\epsilon_{B}$) when modelling high energy emission from astrophysical sources.  There is significant variation of these parameters behind the shock - in particular, $\epsilon_{B}$ decays behind the flow front in the realistic mass ratio case.  This should be accounted for in any detailed modelling of emission from relativistic shocks.

We emphasize that we have only run our simulations for $2500 w_{p}^{-1} = 0.04n_{1}^{-1/2}$ seconds in the frame of the contact discontinuity.  These are short timescales in the context of GRBs (in which a typical pulse lasts from milliseconds to $\sim$ 10 seconds).  However, our primary point is that there is already a clear and growing divergence of results between the realistic mass case and artificial lower mass ratios, in contrast to previous claims that simulations are not sensitive to this ratio.  Hence, we stress the importance of making comparisons among different particle-in-cell codes.  We plan to continue exploring these issues, in addition to extending our simulations to longer times, and adding a particle tracker to discern the details of the acceleration process.
\medskip

\section{Acknowledgements}
 We are very grateful to Bill Daughton for providing access to the VPIC code and for helpful discussions, and to Fan Guo for help with the code to compute particle energy distributions and useful discussions.  We also thank Chengkun Huang for several useful discussions on numerical instabilities in relativistic PIC simulations.  This work is supported in part by the M. Hildred Blewett Fellowship of the American Physical Society, www.aps.org.  Work at LANL was done under the auspices of the National Nuclear Security Administration of the U.S. Department of Energy at Los Alamos National Laboratory, LA-UR-16-21215.

\end{document}